# Characterization of Defect Structure in Electrodeposited Nanocrystalline Ni Films

Tamás Kolonits[a,*], Péter Jenei[b], Bence G. Tóth[c], Zsolt Czigány[a], Jenő Gubicza[b], László Péter[c] and Imre Bakonyi[c]

[a]*Institute for Technical Physics and Materials Science,
Centre for Energy Research, Hungarian Academy of Sciences.
Konkoly-Thege M. út 29-33., H-1121 Budapest, Hungary*

[b]*Eötvös Loránd University, Department of Material Physics.
Pázmány Péter sétány 1/A, H-1117 Budapest, Hungary*

[c]*Wigner Research Centre for Physics, Hungarian Academy of Sciences.
H-1121 Budapest, Konkoly-Thege út 29-33, Hungary*

**Abstract**

The microstructure of electrodeposited Ni films produced without and with organic additives (saccharin and formic acid) was investigated by X-ray diffraction (XRD) line profile analysis and cross-sectional transmission electron microscopy (TEM). Whereas the general effect of these additives on the microstructure (elimination of columnar growth as well as grain refinement) was reproduced, the pronounced intention of this study was to compare the results of various seldom-used high-performance structural characterization methods on identical electrodeposited specimens in order to reveal fine details of structural changes qualitatively not very common in this field. In the film deposited without additives, a columnar structure was observed showing similarities to the T-zone of structure zone models. Both formic acid and saccharin additives resulted in equiaxed grains with reduced size, as well as increased dislocation and twin fault densities in the nanocrystalline films. Moreover, the structure became homogeneous and free of texture within the total film thickness due to the additives. Saccharin yielded smaller grain size and larger defect density than formic acid. A detailed analysis of the grain size and twin boundary spacing distributions was carried out with the complementary application of TEM and XRD, by carefully distinguishing between the TEM and XRD grain sizes.

---

* Corresponding author. E-mail: kolonits@mfa.kfki.hu

# 1. Introduction

The structural features of nanocrystalline materials have a deterministic effect on the macroscopic properties of the material. Grain boundaries and crystal defects increase electron scattering and hinder dislocation propagation, thereby increasing the electrical resistivity and the mechanical hardness of the material, respectively. Therefore, it is essential to determine the fundamental structural parameters to understand the effect of the processing conditions on the materials properties. Both the grain size and defect density can be influenced by the purity of the material, the parameters of the synthesis process and the preparation method itself.

Electrodeposited thin films often have a grain size in the nanocrystalline range [1-3]. Most of the early efforts on electrodeposited nanocrystalline metals were concentrated on nanocrystalline Ni deposits. Reported studies included hardness [2], electrical transport [3] and magnetic [3,4] properties, corrosion behavior [5], hydrogen diffusion [6] and thermal stability [7]. The dependence of grain size on the deposition parameters was investigated by transmission electron microscopy (TEM) and relationships between electrical resistivity and deposition parameters were established [3,8-10]. Since then, the mechanical performance of electrodeposited nanocrystalline Ni films has been extensively studied. Several papers have reported enhanced mechanical properties like increased hardness [2,11-13] or tensile strength [13-18]. The validity of a Hall-Petch [11,13,19] or an inverse Hall-Petch [20] relation at small grain size in tensile and indentation tests of electrodeposited nanocrystalline Ni has been particularly discussed in detail. The contributions of different deformation mechanisms to plasticity, like intragranular dislocation plasticity and grain boundary mediated processes have also been investigated [14,15,21,22]. An increase in corrosion resistance of electrodeposited nanocrystalline nickel compared to coarse-grained counterparts has been reported [5,23-27]. Superhydrophobic nanostructured surfaces which produce the so-called lotus-effect can also be achieved by electrodeposition of Ni [28].

The microstructure and grain size refinement in electrodeposited nickel can be controlled by several deposition parameters like bath temperature [19,29,30], solution pH [16] and applied current density [3,9,18,19,29,31-34]. Generally, the grain size increases with temperature [19,29,30] and it exhibits a minimum as a function of the pH [16] whereas the grain size variation with current density is controversal. In several studies, the grain size was found to decrease with deposition current density [19,29,31,32]. According to these observations, Ni deposition is conformal to the general trends observed for electrodeposited metals [35,36]. However, there have been also several reports where an opposite trend was observed: the grain size increased with applied deposition current density [3,9,18,33,34]. Rashidi and Ahmadeh [31] summarized explanations given previously for this unusual behavior. They noticed that also the different kinetic parameters of the various baths should certainly be accounted for when attempting to interpret the two different characteristics observed.

The grain refinement effect of several additives including formic acid [3,8,9,37], saccharin [23,24,27,29-31,38] and others [12,16,18,19,33,38,39] has also been tested and an average grain size in the range of 10-50 nm could be achieved. Saccharin has long been extensively used as grain refiner, being one of the most common additives in industrial nickel baths and is known as stress reliever and hardener [39]. It was found [29,30] that up to a certain concentration, the amount of saccharin in the bath decreases the grain size and for higher additive concentrations, the grain size saturates. The saccharin concentration that leads to the saturation of the grain size varies from work to work. It is common that an exponential-like decrease in the grain size is found [29,30,40] with an at least three-fold decrease in grain size at the saturation level as compared to the additive-free solution. The big majority of this grain size refinement takes place already at 1 gram/liter saccharin concentration for pure Ni deposits [29,30,40,41]. On the other hand, the grain size dependence of the Ni deposits produced in the presence of formic acid has not been studied as a function of the additive concentration. For Ni-Zn deposits, it was found that the deposit properties studied saturated already at 0.1 mol/liter formic acid concentration [42].

It was also shown that for a given set of pulse-plating deposition conditions from an additive-free Watts type bath the grain size could only be reduced down to about 100 nm [40]. However, the same pulse-plating parameters resulted in grain sizes down to 10 nm when saccharin was also added to the Watts bath [43]. A small amount of S and C was found to be incorporated in the deposit from the saccharin-containing bath [43] and the presence of these elements in the deposit could have been a reason for the smaller grain sizes. However, the grain refinement was rather ascribed [40] to the presence of the saccharin additive in the bath which acts as an inhibitor in a sense that the saccharin molecules adsorbed on the deposit surface hinder the surface mobility of Ni-containing reaction intermediates. As a result, the growth of existing nuclei on the deposit surface may be blocked and this leads to more frequent nucleation than in the absence of the additive in the bath. It is generally accepted [39] that bath additives have an effect on the growth mode and, thus, on the defect structure including the grain size.

However, the formation of defects and the relative importance of their types, i.e., dislocations, twin faults and their density as an effect of additives have not been investigated in detail yet. Therefore, the purpose of this work is to present the results of a comprehensive study on the microstructural effects of two organic additives: formic acid and saccharin, also by comparing the results of two complementary structural investigation methods on identical specimens.

An effective experimental technique for the characterization of the crystallite size and defect structure in nanocrystalline materials is X-ray diffraction (XRD) line profile analysis [44,45]. XRD line profile analysis provides a rich and reliable statistics about crystallite size, dislocation density and twin fault probability due to the large irradiated volume. Line profile evaluation procedures are indirect methods: we have a number of presumptions about the microstructure (e.g., distribution of twin fault spacing, shape of the crystallite or the type of the crystallite size distribution function) and we calculate the parameters of this model.

Another method for the characterization of grain size and defect structure in nanocrystalline materials is transmission electron microscopy (TEM) [46] which provides local information with less statistics. However, TEM is a direct procedure for the study of microstructure; therefore, it is capable of answering questions like detailed grain size and defect spacing distributions, as well as their homogeneity (if there are any gradient or fluctuations of the parameters in the direction perpendicular to the film growth direction), (an)isotropy and spatial variation along the film growth direction. Therefore, TEM is an effective complementary technique to XRD line profile analysis, being informative regarding the assumptions of the profile fitting methods for the grain size and twin fault spacing distributions (e.g. lognormal and geometric distributions, respectively, in case of the method used in this work).

The grain size directly determined by TEM often differs from the crystallite size obtained by XRD line profile analysis [45,47,48]. The latter quantity always gives the size of domains which scatter X-rays coherently. The coherency of X-ray photons can be broken by lattice defects inside the grains [47]. For instance, dislocation walls and subgrain boundaries may yield small misorientations (lower than 1-2 degrees), leading to incoherent scattering from the separated domains. It has been shown that the break of coherency can occur even for dipolar dislocation walls, although they do not cause any misorientations between the domains [47]. Since the lattice defects fragment the grains into smaller coherently scattering domains, the crystallite size determined by XRD line profile analysis is usually smaller than the grain size obtained by TEM [48]. This difference depends on the processing method of the studied material. For instance, in the case of severely deformed metallic materials, the TEM grain size may be larger by a factor of six than the XRD crystallite size due to the large density of dislocations formed inside the grains during deformation [45]. For metallic materials produced without severe plastic deformation, this deviation is usually smaller [45] and for ceramic materials even a coincidence between the grain and crystallite sizes can be observed [49].

By keeping in mind the above described distinction between grain size and crystallite size, the effect of additives on the grain/crystallite size and on the defect density in electrodeposited Ni films was investigated by combined XRD line profile analysis and TEM along this line. The amount of eventually incorporated elements of the additives in the deposit was measured by energy-dispersive X-ray spectroscopy (EDS). Besides the characterization of the defect structure, the results obtained from the two complementary techniques of XRD and TEM were compared and the assumptions of the XRD line profile analysis method were confirmed by TEM. An important feature of the present work is the cross-sectional TEM study since most previous TEM investigations on electrodeposited Ni, except Ref. 34, have been carried out in plane-view mode only.

## 2. Experimental

*2.1. Processing of Ni by electrodeposition*

For the deposition of the samples, three aqueous electrolytes were used which were prepared with chemicals from Reanal (except where otherwise indicated). There was no additive in the first electrolyte (referred to as "NA") and its composition was 0.6 mol/$\ell$ nickel sulfate (from $NiSO_4 \cdot 7 H_2O$, Alfa Aesar, 98.0 %), 0.30 mol/$\ell$ sodium sulfate (from $Na_2SO_4 \cdot 10 H_2O$), 0.25 mol/$\ell$ $H_3BO_3$ and 0.15 mol/$\ell$ $H_3NO_3S$ [50]. The composition of the second electrolyte was the same as that of the "NA" bath except for the addition of 46 m$\ell$/$\ell$ (equal to 1.22 mol/$\ell$) formic acid [9] (called "FA" bath). The composition of third electrolyte was again the same as that of the "NA" bath except for the addition of 1 g/$\ell$ saccharin (called "SC" bath). The Ni samples prepared from the corresponding electrolytes will be denoted as e-Ni-NA, e-Ni-FA and e-Ni-SC, respectively. The pH of all the three solutions was set to 3.25 with NaOH.

The Ni layers with a thickness of about 20 μm were deposited on a thick Ti sheet from which the deposit was peeled off mechanically (with the help of a scotch tape). The deposition was performed in a tubular cell of 8 mm x 20 mm cross section with an upward facing cathode at the bottom of the cell in order to ensure the lateral homogeneity of the deposition conditions and to avoid edge effects [51,52].

Electrodeposition was carried out with direct current at room temperature. The deposition current density was $j = -43.75$ mA cm$^{-2}$. The thickness of the layers was controlled with the deposition time through Faraday's law by assuming 96% current efficiency [50]. Since the compositional and structural studies were performed on both sides of the deposits, the phrase

"electrolyte side" is used to describe the side of the sample towards the solution, while the other specimen side is referred to as "substrate side" and they will be abbreviated as "es" and "ss", respectively. The deposition conditions of the three samples are summarized in Table 1.

*2.2. XRD line profile analysis*

The XRD line profiles were measured by a high-resolution rotating anode diffractometer (type: RA-MultiMax9, manufacturer: Rigaku) using CuK$_{\alpha 1}$ radiation with a wavelength of $\lambda = 0.15406$ nm. Two-dimensional imaging plates detected the Debye-Scherrer diffraction rings. The line profiles were determined as the intensity distribution perpendicular to the rings and obtained by integrating the two-dimensional intensity distribution along the rings. The line profiles were evaluated by the extended Convolutional Multiple Whole Profile (eCMWP) fitting procedure [53,54]. In this method, the diffraction pattern is fitted by the sum of a background spline and the theoretical peak profiles. The latter ones were obtained as the convolution of the instrumental pattern and the theoretical line profiles related to the crystallite size, dislocations and twin faults. As an example, the fitting for the film deposited with saccharin additive is shown in Fig. 1. The details of the eCMWP procedure can be found in Refs. 53 and 54. Because of the nano-grained microstructure of the studied samples, the physical broadening of the profiles was much higher than the instrumental broadening. Therefore, instrumental correction was not applied in the evaluation. This method gives the median ($m$) and the lognormal variance ($\sigma^2$) of the assumed lognormal crystallite size distribution, the dislocation density ($\rho$) and the twin fault probability ($\beta$) with good statistics, where the twin fault probability is defined as the fraction of twin boundaries among the (111) lattice planes. The arithmetically averaged crystallite size can be obtained as: $<x>_{\text{arit}} = m \times \exp(0.5\,\sigma^2)$.

*2.3. TEM investigations*

The microstructure of the films was also investigated by TEM using two different instruments:

a Philips CM20 transmission electron microscope operated at an acceleration voltage of 200 kV, while high-resolution TEM (HRTEM) investigations were carried out by a JEOL JM3010 transmission electron microscope operated at an acceleration voltage of 300 kV with a 0.17 nm point resolution. TEM images were taken in cross-sectional view. Cross-sectional samples were prepared by creating a "sandwich" from two 1.8 x 0.5 mm$^2$ pieces of the sample (film to film) placed between two slices of Si single crystal with the same dimensions. This "sandwich" was mounted and glued (with an araldite-based conductive glue) into the window of a Ti grid. A heat treatment at 160-180 °C was followed by mechanical thinning, polishing, and dimpling to a thickness of ca. 20 μm in the middle. Thinning to electron transparency was achieved by ion-beam milling [55] using a Technoorg Linda ionmill with 10 keV Ar$^+$ ions at an incidence angle of 5° with respect to the surface. In the final period of the milling process, the ion energy was decreased gradually to 3 keV to minimize ion-induced structural changes in the surface layers. The estimated temperature of the sample was not more than 180 °C, therefore it was supposed that the microstructure of the sample has not changed during TEM sample preparation [7,56].

To achieve good balance between statistics and resolution (and the minimal measurement error of distances) bright-field (BF) and dark-field (DF) images were taken with a magnification of 50.000 x. To confirm the direction independence of the twin boundaries, the samples were tilted in an angular region ±12°.

Both BF and DF images were taken by TEM. In the case of bright-field images, more grains in the sample are visible but their size can be determined with larger uncertainty. In contrast, dark-field images [46] show only those grains of the sample which reflect the electron beam in the same specific direction (therefore, having the same orientation). Thus, these images show a relatively small number of grains, but their size can be determined with higher accuracy. Therefore, for the measurement of the grain size, one bright-field and several dark-field images (with different selected orientations) were taken in the same area of the sample. To achieve good statistics, approximately 1000 grain sizes were extracted from BF and DF images for each samples.

*2.4. Composition analysis*

In order to determine the deposit composition, energy-dispersive X-ray spectroscopy (EDS) measurements were performed by a JEOL JSM-25S scanning electron microscope equipped with a RÖNTEC EDS System with an UHV Dewar Si(Li) detector (Röntec GmbH). In the samples, volumes with the dimensions of ~8000 $\mu m^2 \times 2$ μm were excited using an electron beam with an energy of 25 keV. Both the substrate and electrolyte sides of the samples were examined.

**4. Results and discussion**

*4.1 TEM grain sizes and XRD crystallite sizes of electrodeposited Ni films*

Figure 2 shows TEM images, together with the corresponding selected-area electron diffraction (SAED) patterns, obtained on the cross-section of a Ni layer (sample e-Ni-NA) deposited without additives at the electrolyte side (upper panels) and at the substrate side (lower panels). In this sample, a crystallographic texture was detected by XRD where the (220) planes are parallel to the film surface. Similar texture was also observed in electrodeposited nanocrystalline Ni [57,58] and Ni-Co alloy [50] deposits. According to Fig. 2, the grains are nearly equiaxed at the substrate side of the film and are smaller compared to those at the electrolyte side. At the substrate side, the average crytallite size is ~50 nm. The film growth gradually transforms to a columnar grain morphology with more than 3 μm long and ~120 nm wide columns. The latter dimension is determined from the TEM study; hereafter, this value will be regarded as grain size. The structural modification along the growth direction from disordered small crystals to a columnar system is common for the electrodeposition process in the case of weakly inhibited (additive-free) baths and direct-current deposition [33,39].

The described film morphology is typical for T-zone of structure zone models [59] developed for physical vapor deposition (PVD) film growth where the substrate temperature is between 0.1 and 0.3 times the melting point of the deposited material. In this temperature range, the grain growth is hindered in the bulk of the film (because of the low speed of bulk diffusion) but the surface morphology supports competitive growth of grains and selection of grain orientations resulting in the formation of texture [59]. During electrodeposition, the substrate temperature is similarly low to hinder diffusion process in the bulk, but there is sufficient surface mobility at the deposit–electrolyte interface for the development of a competitive growth.

Figure 3 shows cross-sectional TEM images and corresponding SAED patterns at the electrolyte side of Ni films deposited with the addition of formic acid or saccharin. In contrast to the film deposited with no organic additives (Fig. 2), for the films produced with additives, no difference between the microstructures at the substrate and electrolyte sides was observed. The grains are rather equiaxed throughout the whole thickness.

The size distributions of grains determined from the TEM images for the Ni deposits obtained with the additives formic acid and saccharin are plotted as histograms in Fig. 4 and Fig. 5, respectively. The grain size distributions can be well fitted by lognormal distributions (referring to the assumptions of the current XRD line profile analysis model described in Section 2.2). The parameters of the lognormal size distributions (median and lognormal variance) fitted to the data are listed in Table 2, together with results for the e-Ni-NA deposit for which these data were determined for both sides of the deposit. The additives resulted in a decrease in grain size and the effect was stronger for saccharin. Since the additive concentrations applied in the present work resulted in the saturation of the grain size (see the discussion on this issue in the Introduction), it can be claimed that saccharin is a stronger grain refinement agent than formic acid.

The two parameters of the lognormal crystallite size distribution and the average crystallite size determined by XRD peak profile analysis for the three Ni films are also listed in Table 2. In the case of the additive-free sample, the crystallite size is smaller at the substrate side than at the electrolyte side in accordance with the TEM observations. It should be noted that the difference between the XRD crystallite sizes at the two sides of the film e-Ni-NA is smaller than the difference of their TEM grain sizes. This effect can be attributed to the high penetration depth of X-rays (estimated as

~14 μm) compared to the thickness of the film (about 25 μm) since the cross-sectional TEM simages were taken very close to the deposit surfaces on both sides.

According to the data in Table 2, both additives resulted in a reduction of the XRD crystallite size, similarly to the TEM grain size. At the same time, the grain size provided by TEM is larger by a factor of 1.5-2.3 than the XRD crystallite size for the investigated samples. As discussed in the Introduction, this difference is a general phenomenon and can be observed in various materials [60-62] since it is caused by the lattice defects inside the grains (e.g., small-angle grain boundaries not revealed by TEM) which break the coherency of the X-ray radiation. Therefore, the XRD crystallite size rather reflects the size of subgrains in the microstructure. The existence of such subgrains inside the larger grains is illustrated in Fig. 6 by the TEM images obtained on the electrolyte side of the additive-free Ni film. The small areas with different contrasts in both the BF and DF images correspond to subgrains with much smaller size than either the width or the height of the columnar grains. The difference between the grain and crystallite size distributions for a given deposit can also be assessed in Figs. 4 and 5.

*4.2. Dislocation density*

The dislocation density values obtained by the XRD peak profile analysis are listed in Table 3. For the film deposited without additives, the dislocation density is larger at the substrate side than at the electrolyte side. The higher dislocation density at the substrate side is in accordance with the smaller crystallite size since the subgrain boundaries with low misorientations are usually associated with a series of edge dislocations. The additives resulted in a strong increase in the dislocation density. Most probably, this effect is related to the refinement of the microstructure occurred due to additives since in nanocrystalline materials dislocations are not stored in grain/subgrain interiors [63] and the density of dislocations in grain boundaries increases with decreasing grain size. Saccharin is more effective than formic acid in the increase of the dislocation density.

*4.3. Twin-boundary spacing*

Table 4 lists the twin-fault probability ($\beta$) values determined by XRD peak profile analysis. For the sake of an easier comparison of XRD and TEM results, the average twin-boundary spacing was calculated from the twin-fault probability as $d_T = 100 \times d_{111}/\beta$, where $d_{111}$ is the interplanar spacing of (111) lattice planes. In the case of Ni, $d_{111} = 0.2034$ nm. The twin-fault probability was under the detection limit of XRD peak profile analysis on both the substrate and electrolyte sides of the additive-free Ni film. This limit is ~0.1 % which corresponds to a twin-boundary spacing of about 200 nm. The addition of either formic acid or saccharin yields considerable twin-boundary probability ($(0.8\pm0.1)\%$ and $(3.3\pm0.1)\%$ respectively). Saccharin results in a siginificantly higher value of $\beta$ than formic acid (see Table 3).

The twin-boundary spacing was also determined from the TEM images by measuring the twin boundary distances, similarly as was in the case of determining grain sizes. In the additive-free sample, twin boundaries were not observed inside the grains. At the same time, both additives resulted in twinning in the grain interiors. Figures 7 and 8 show BF and DF images obtained on grains containing twin boundaries in the films deposited with formic acid and saccharin, respectively. The twin-boundary spacing distributions obtained by TEM for these two materials are plotted in Figs. 9 and 10. These figures also show the twin-boundary spacing distributions determined by XRD. In the XRD line profile fitting method, it is assumed that the appearance of a twin fault in a lattice plane is stochastic (random twinning) with the probability of $\beta$, i.e., it does not depend on the occurrence of twin boundaries in the neighboring (111) planes. If the crystal between two adjacent twin boundaries is referred to as twin lamella, the probability of the formation of a lamella with the thickness $d_T$, $W(d_T)$, can be obtained as

$$W(d_T) = (1-\beta)^{d_T/d_{111}-1} \beta \quad . \tag{1}$$

Therefore, in this model, the variation of twin-boundary spacing can be described by a geometric distribution density function.

The twin-fault spacing distributions obtained by XRD follow the trends in the histograms determined by TEM, except for spacings smaller than 5 nm. This deviation can be attributed to nanotwins which have a considerable contribution to XRD peak profile broadening; however, they can be hardly observed by TEM. Figure 11 shows an HRTEM image which illustrates the presence of nanotwins in the film deposited with saccharin. The thickness of these twin lamellae varies between 2 and 6 nm. It should be noted, however, that the agreement between the twin-boundary spacings measured by the two methods is still satisfactory, especially if we consider that the volume studied by XRD is at least six orders of magnitude larger than that examined by TEM. Anisotropy of the orientation of twin boundaries and grain boundaries was not detected either by TEM or by XRD.

*4.4 Effect of additives on the microstructure and defect formation*

Generally, in electrodeposition of metals, the application of organic additives results in smaller coherently scattering domains and higher density of dislocations, twins and stacking faults; however, the exact details of the mechanism of these effects is still under debate [64]. According to our EDS results, the deposited films contain a detectable amount of Co and Fe impurities. These metals are known to deposit besides Ni in an anomalous way, i.e., their concentration is higher in the deposit than in the precursor chemical [65,66]. Formic acid and saccharin additives did not influence the degree of Fe and Co enrichment, therefore these metallic elements cannot cause the change in grain size and lattice defect densities. The only significant difference between the chemical compositions of the three kinds of film is that ~0.3 at.% sulphur was detected in the film with saccharin additive.

According to former studies, the common effect of all additives (among others, formic acid and saccharin) is the following [67]: hydrophilic functional groups of organic additives (e.g., $SO_2$ or carboxyl group) can enclave the additive molecule between water (in the electrolyte) and the previously deposited metal surface. This "organic shield" does not block the electron transfer but hinders the ion transfer. The preferential sites for the adsorption of the surface-active additive molecules are the kink sites and step edges. Since these crystal features are, at the same time, also the most active growth centers, their blocking leads to the necessity of new grain nucleation, which obviously results in grain refinement. This shielding mechanism on the growth surface can inhibit the grain growth and induce secondary nucleation (i.e., nucleation taking place not on the foreign substrate but on the growing deposit). Having no epitaxial relation between the newly nucleated grains and the grains beneath, the shielding mechanism inhibits the development of texture as well. It is noted that when using formic acid as additive, components from the inhibiting additive could not be detected in the deposit by EDS analysis; however, impurity segregation at grain boundaries below the EDS detection limit cannot be excluded. Such a small impurity level may have significant effect on the grain size as pointed out formerly [59,64]. According to Refs. 30 and 68, in the case of saccharin, passivation of the cathode surface by precipitation of nickel hydroxide gives rise to the nucleation of new grains. Most probably, this electrochemical process is enhanced by the segregation of sulphur or sulphide phases that can occur both within the grains and at the grain boundaries [64]. Similar explanation for grain size refinement by the segregation of an oxide phase was established for PVD films [59].

Beside the hindrance of the characteristic crystal growth sites, saccharine also decomposes upon the electrodeposition of Ni, resulting in benzamide formation and sulfur inclusion [69]:

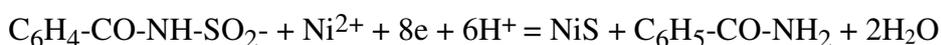
$C_6H_4\text{-}CO\text{-}NH\text{-}SO_2\text{-} + Ni^{2+} + 8e + 6H^+ = NiS + C_6H_5\text{-}CO\text{-}NH_2 + 2H_2O$

Although the formation of a well-defined NiS compound may not take place, the inclusion of the sulfur atom in the Ni lattice leads to a lattice deformation that impedes the coherent growth of the crystals and to a higher secondary nucleation probability. Hence, the sulfur inclusion has a contribution to the grain refinement. According to the above reaction pathway, the enhancement of carbon inclusion is not a necessary consequence of the sulfur inclusion [69].

The formation of twins at the growth surface may also be facilitated by impurity segregation on the close-packed (111) crystal facets. The segregated species can serve as 2D nucleation centers for the successive (111) plane; however they may cause a shift of the plane compared to the ideal fcc stacking, resulting in twin boundaries or stacking faults. In addition, the elastic stresses developed inside the grains during deposition can relax by forming lattice defects such as twin boundaries.

The dislocation density also increases due to the addition of formic acid and saccharin. Most probably, this effect is associated with the refinement of the microstructure owing to additives. As the interiors of nanograins usually do not store dislocations [63], their majority is assumed to be located in the grain/subgrain boundaries. These dislocations may form in order to reduce the elastic incompatibility stresses developed between the grains/subgrains at their boundaries.

## 5. Conclusions

The microstructure of electrodeposited Ni films produced without and with organic additives (saccharin and formic acid) was investigated by X-ray diffraction (XRD) line profile analysis and cross-sectional transmission electron microscopy (TEM). The well-known effect of these additives on the microstructure (elimination of texture and columnar growth as well as grain refinement) was reproduced. However, the pronounced intention of this study was to compare the results of various seldom-used high-performance structural characterization methods on identical electrodeposited specimens in order to reveal fine details of structural changes qualitatively not very common in this field. A detailed analysis of the grain size and twin boundary spacing distributions was carried out with the complementary application of TEM and XRD. Special emphasis was put on distinguishing between the grain size as derived directly from TEM studies and the size of the coherently scattering regions (termed as crystallite size) revealed by XRD.

The following results were obtained:

1. In the film deposited without additives (e-Ni-NA), mainly a columnar grain structure was observed towards the solution side of the film, showing similarities to the T-zone of structure zone models [59]. The height of the columns was larger than 3 µm, while their width was ~120 nm. In this sample, a crystallographic texture was also detected where the planes (220) are parallel to the film surface. Close to the substrate, the film exhibited smaller and equiaxed grains. Considerable twinning was not observed either by XRD or TEM.

2. Both additives eliminated the texture and the columnar grain structure, yielding fine-grained isotropic microstructures throughout the total thickness, with grain sizes of about 48 nm and 24 nm for formic acid and saccharin, respectively, as determined by TEM. Both XRD and TEM methods evidenced that the additives increase the density of lattice defects (dislocations and twin faults). The additive saccharin was more effective in the refinement of the grain size and the enhancement of lattice defect densities when compared to formic acid. The hindering effect of both additives on the grain growth during deposition can be attributed to the "shields" formed from the hydrophilic functional groups of organic additives [67]. A significant amount of sulphur (~0.3 *at.%*) was detected in the film deposited with saccharin, indicating the possibility of different mechanisms for the grain size refinement for the two additives. In case of formic acid, no detectable incorporation of inhibiting additives could be revealed by EDS analysis.

3. The tendency in the additive-induced grain size evolution determined by TEM was similar to the trend in the crystallite size obtained by XRD line profile analysis. Presumptions about the grain size distribution (lognormal) of the eCMWP model were confirmed. However, the grain size values determined by TEM are higher by a factor of 1.5-2.3 than the crystallite sizes determined by XRD line profile analysis. This difference can be explained by the fragmentation of larger grains into smaller domains (subgrains) due to the various lattice defects within the grain interior.

4. TEM revealed that the distribution of twin-boundary spacing follows geometric distribution for large spacing values in accordance with the model used in the XRD line profile analysis. However, for twin-boundary spacings smaller than 5 nm, TEM yields a smaller fraction than that predicted by XRD. HRTEM evidences the presence of numerous nanotwins, suggesting that the difference between the twin boundary densities determined by TEM and XRD line profile analysis is caused by the difficulty in the observation of nanotwins with good statistics by TEM.

**Acknowledgement**

Dr. Attila Tóth and dr. Levente Illés are acknowledged for the assistance in the EDS measurements. This work was supported by the Hungarian Scientific Research Fund (OTKA) through grants No. K-109021 and K-104696.

# Tables

| Sample | Electrolyte/Additive | Substrate | $j$ [mA cm$^{-2}$] | Thickness |
|---|---|---|---|---|
| e-Ni-NA | No additive | Ti | 43.75 | 24±3 μm |
| e-Ni-FA | Formic acid (46 m$\ell/\ell$) | Ti | 43.75 | 30±5 μm |
| e-Ni-SC | Saccharin (1 g/$\ell$) | Ti | 43.75 | 29±3 μm |

**Table 1:** Deposition parameters of samples investigated in the present work. The composition of the electrolytes is described in section 2.1.

| Sample | TEM grain size | | | XRD crystallite size | | |
|---|---|---|---|---|---|---|
| | $<x>$arit [nm] | $\sigma$ | $m$ [nm] | $<x>$arit [nm] | $\sigma$ | $m$ [nm] |
| e-Ni-NA (es) | 124±1 | 0.57±0.01 | 105±2 | 55±6 | 0.06±0.02 | 55±4 |
| e-Ni-NA (ss) | 50±1 | 0.71±0.01 | 39±1 | 34±4 | 0.40±0.13 | 31±5 |
| e-Ni-FA | 48±1 | 0.71±0.03 | 37±2 | 23±3 | 0.40±0.03 | 21±2 |
| e-Ni-SC | 24±1 | 0.67±0.05 | 19±1 | 16±2 | 0.47±0.04 | 14±1 |

**Table 2:** Comparison of TEM grain sizes and XRD crystallite sizes for electrodeposited nanocrystalline Ni samples prepared with different additives: e-Ni-NA: no additive; e-Ni-FA: formic acid; e-Ni-SC: saccharin. Due to the heterogeneity of sample e-Ni-NA along the growth direction, the substrate (ss) and electrolyte/solution (es) sides were distinguished. At the electrolyte side of sample e-Ni-NA, there is a columnar grain structure and the grain size specified corresponds to the width of the columns; the length of the columns is larger than 3 μm.

| Sample | $\rho$ [10$^{14}$ m$^{-2}$] |
|---|---|
| e-Ni-NA(es) | 9 ± 1 |
| e-Ni-NA(ss) | 16 ± 2 |
| e-Ni-FA | 46 ± 5 |
| e-Ni-SC | 176 ± 18 |

**Table 3:** Dislocation densities obtained by XRD profile analysis.

| Sample | $\beta$ [%] | $d_{T,XRD}$ [nm] | $d_{T,TEM}$ [nm] |
|---|---|---|---|
| e-Ni-NA | 0 ± 0.1 | n.a. | n.a. |
| e-Ni-FA | 0.8 ± 0.1 | 25 ± 3 | 25.1 ± 0.1 |
| e-Ni-SC | 3.3 ± 0.1 | 6.1 ± 0.5 | 17.3 ± 0.1 |

**Table 4:** Twin-fault probability ($\beta$) determined by XRD profile analysis. The average twin-boundary spacing ($d_{T,XRD}$) determined from $\beta$ is also listed. For comparison, the average twin-boundary spacing obtained by TEM ($d_{T,TEM}$) is also provided.

# Figures

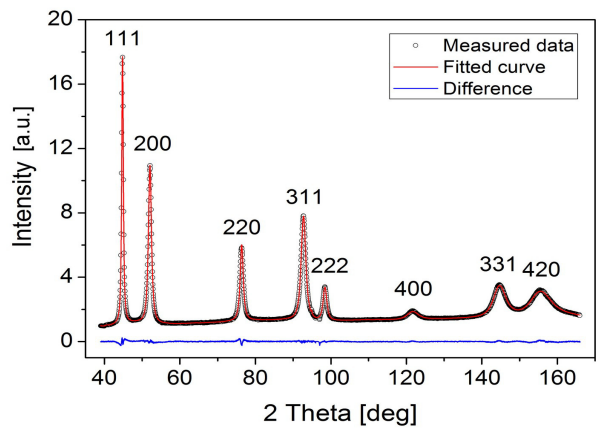

**Fig. 1** XRD pattern of sample e-Ni-SC. The experimental diffractogram was fitted by the eCMWP procedure.

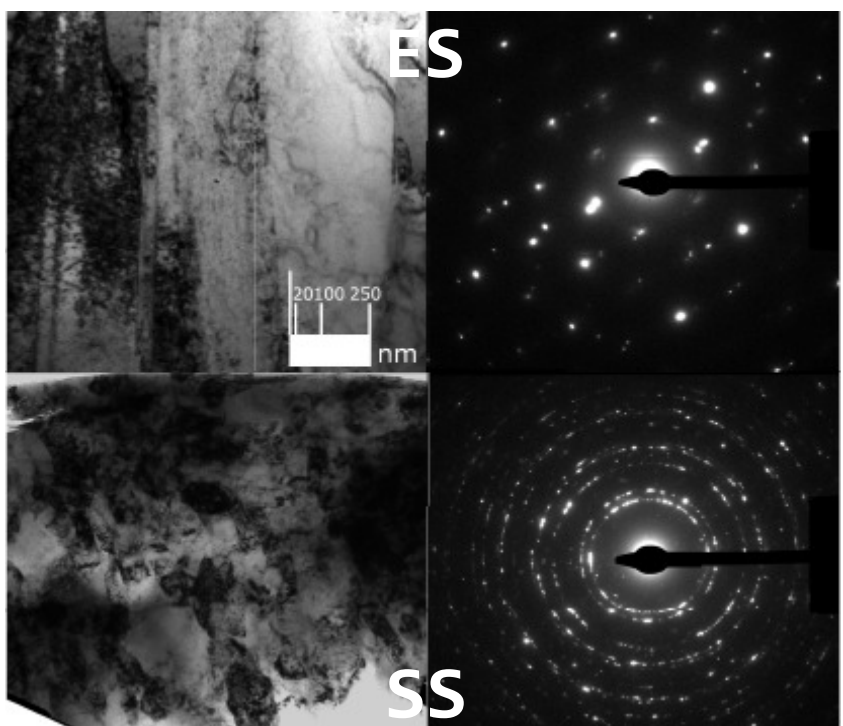

**Fig. 2** Bright-field TEM images and corresponding SAED patterns of the sample with no additive (e-Ni-NA); top: electrolyte/solution side (ES); bottom: substrate side (SS).

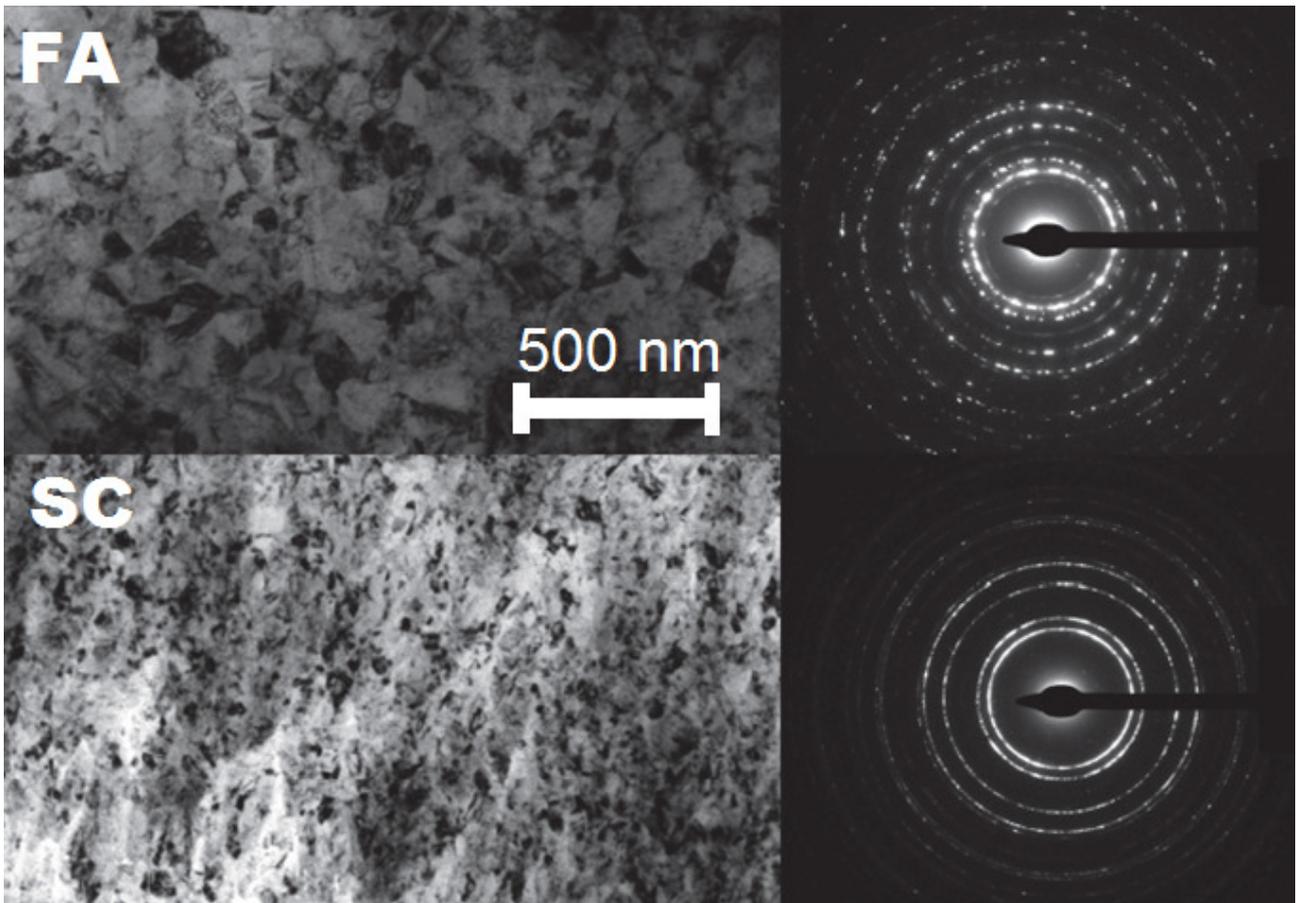

**Fig. 3** Bright-field TEM images and corresponding SAED patterns of samples e-Ni-FA (top) and e-Ni-SC (bottom). The images were taken at the electrolyte/solution side of the corresponding sample. Very similar TEM images and SAED patterns were obtained also at the substrate side of each samples.

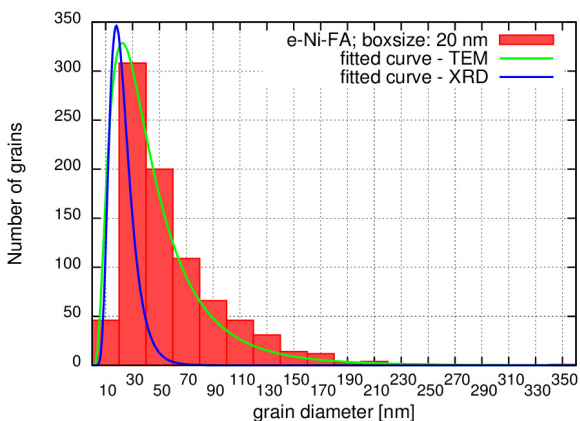

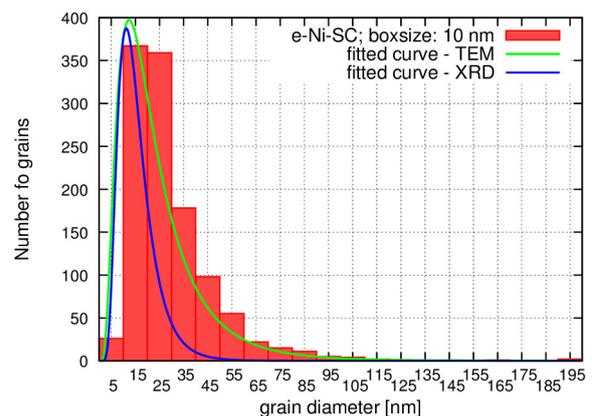

**Fig. 4** Grain size distribution in the film deposited with formic acid as additive (sample e-Ni-FA). Green curve: log-normal grain size distribution function fitted to the histogram data obtained by TEM. Blue curve: log-normal crystallite size distribution function calculated from XRD line profiles. The TEM histogram was obtained by the evaluation of 841 grains.

**Fig. 5** Grain size distribution in the film deposited with saccharin as additive (sample e-Ni-SC). Green curve: log-normal grain size distribution function fitted to histogram data obtained by TEM. Blue curve: log-normal crystallite size distribution function calculated from XRD line profiles. The TEM histogram was obtained by the evaluation of 1145 grains.

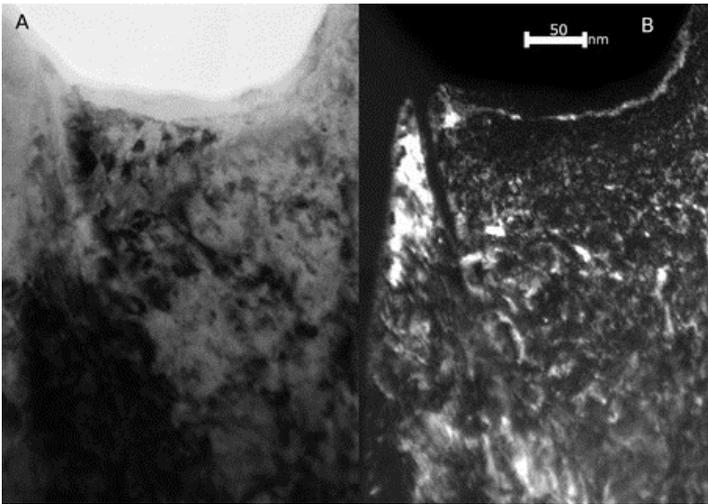

**Fig. 6** Bright-field (A, left) and dark-field (B, right) TEM images obtained at identical magnification on the same area at the electrolyte side of sample e-Ni-NA deposited without additive. The small areas with different contrasts in the BF and DF images correspond to subgrains.

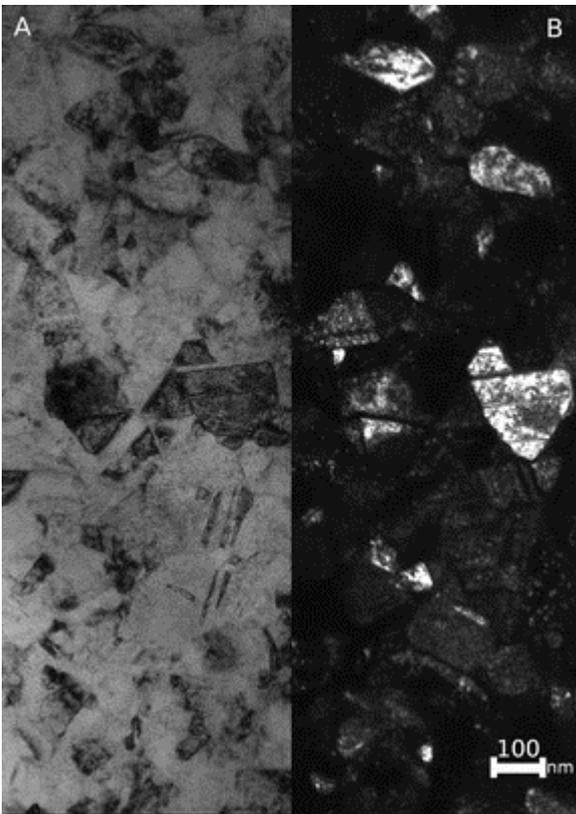

**Fig. 7** Bright-field (left, A) and dark-field (right, B) TEM images taken at identical magnification on the same area of sample e-Ni-FA deposited with additive formic acid.

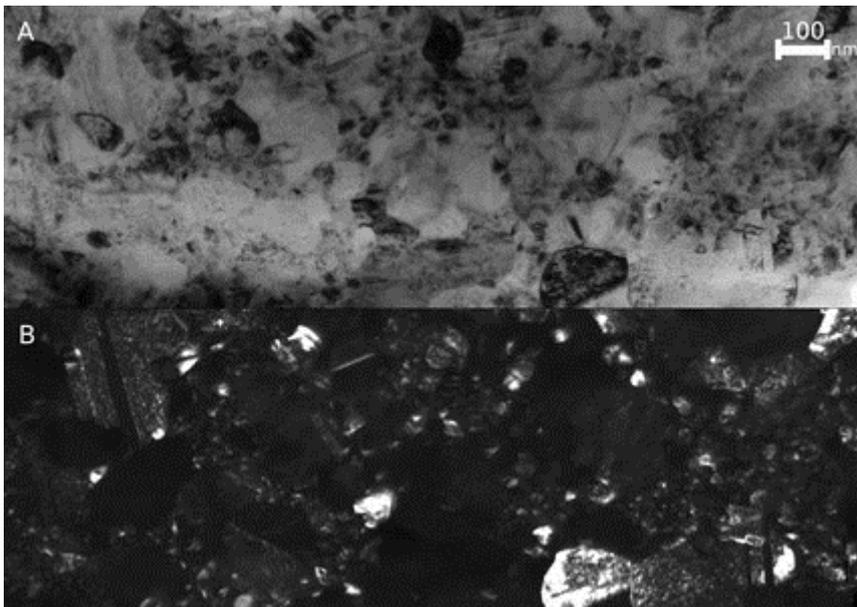

**Fig. 8** Bright-field (top, A) and dark-field (bottom, B) TEM images at identical magnification on the same area of sample e-Ni-SC deposited with saccharin.

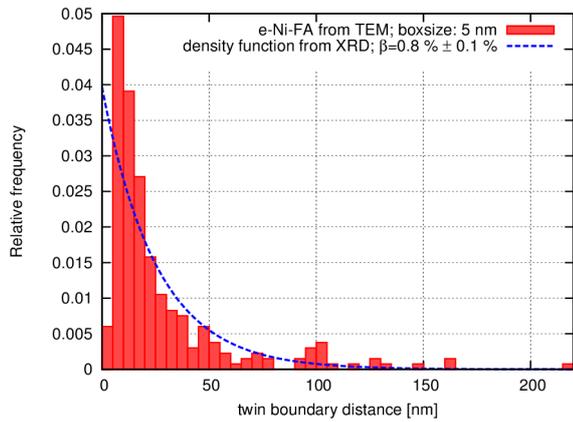

**Fig. 9** Twin boundary spacing distribution function for the film deposited with formic acid as additive (sample e-Ni-FA). The histogram was obtained from TEM images, and the blue line was determined by XRD measurements.

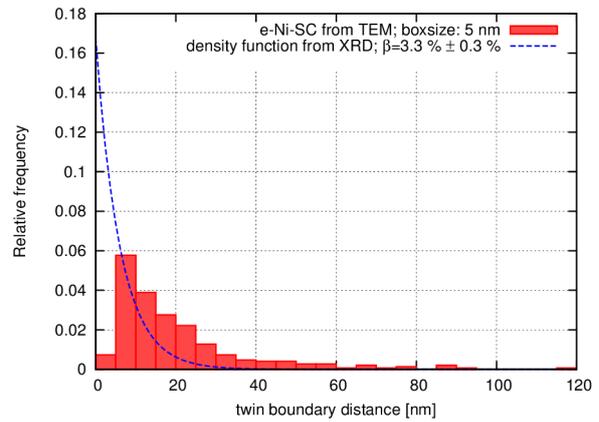

**Fig. 10** Twin boundary spacing distribution function for the film deposited with saccharin as additive (sample e-Ni-SC). The histogram was obtained from TEM images, and the blue line was determined by XRD measurements.

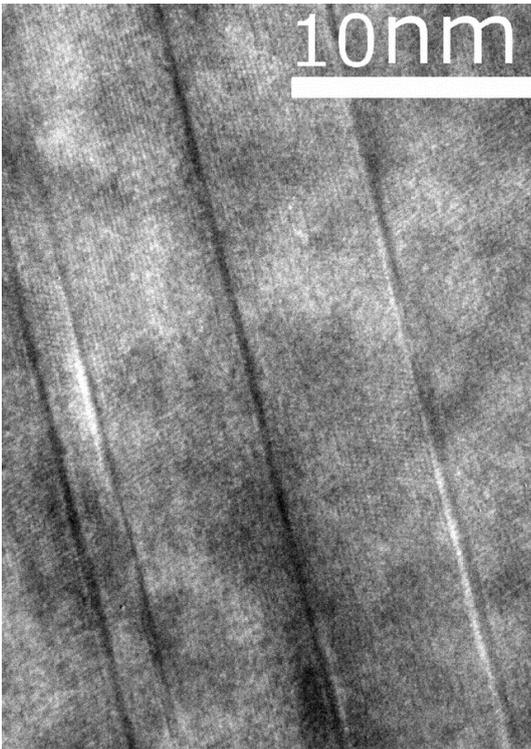

**Fig. 11** HRTEM image (from <110> zone axis) of nanotwin(s) of sample e-Ni-SC obtained with saccharin additive. The distances between the twin boundaries are 2 nm, 6 nm and 6 nm.